\documentclass[11pt]{article}
\usepackage{epsfig}
\usepackage[english,english]{babel}

\begin{document}

%
%
\title{A perturbative moment approach to option pricing.}
\author{Marco Airoldi \\
	Mediobanca, \\
        Piazzetta Enrico Cuccia 1, Milan, ITALY 
	\\
        E--mail: Marco.Airoldi@mediobanca.it
} 

\date{\today}

\maketitle


%
%
\begin{abstract}

In this paper we present a new methodology for option pricing. The main 
idea consists to represent a generic probability distribution function 
(PDF) via a perturbative expansion around a given, simpler, PDF (typically 
a gaussian function) by matching moments of increasing order. Because, as 
shown in literature, the pricing of path dependent European options can 
be often reduced to recursive (or nested) one-dimensional integral 
calculations, the above perturbative moment expansion (PME) leads very 
quickly to excellent numerical solutions. \\
In this paper, we present the basic ideas of the method and the relative
applications to a variety of contracts, mainly: asian,  reverse cliquet 
and barrier options. A comparison with other numerical techniques is also 
presented.  
 
\end{abstract}

%


%
%
%


%
%
\section{Introduction} \label{section:introduction}


A central problem in today finance is to price correctly and rapidly 
exotic options. 
Indeed, in the last years, exotic options have became quite popular and 
financial institutions traded larger and larger quantity of these 
sophisticated financial instruments. Moreover, it has became quite 
common to encounter exotic options embedded in structured bonds, which 
are addressed also to small investors. 
There are no limits to market fantasy in the construction of new 
option varieties. It is therefore of great importance to develop secure 
and fast methodologies for option pricing. The literature on this argument 
is quite huge. Many approaches have been proposed, the most important 
of which are summarized below: 

\begin{itemize}
	\item[(1)] 
	{\bf analytical solutions}: usually they are limited to some special 
	cases (simple contracts where the underlying asset is usually 
	assumed to be governed by a log-normal process)~\cite{Black}.
	Furthermore these exact solutions cannot be extended to more 
	complex contracts. 
	\item[(2)]  
	{\bf binomial or trinomial tree}: these popular approaches are 
	extensively used both in academic literature~\cite{Hull_tree, 
	Barraquand_tree} as well as financial institutions because they 
	can be easily adapted to many different situations and are simple 
	from a conceptual point of view. However many drawbacks are 
	present: 
	1) the convergence is generally slow and the algorithm becomes 
	rapidly inefficient in high dimension (e.g. for barrier options, 
	the problem dimension scales linearly with the number of monitoring 
	dates); 
	2) the error is not monotonic by increasing the number of 
	nodes~\cite{Boyle_tree}; 
	3) usually the underlying movements are assumed to be described by 
	a log-normal process; 4) the result accuracy depends on how the 
	lattice is defined and last but not least, the method becomes 
	critical for particular choice of market data (e.g. for barrier 
	options, when the spot price is close to barrier 
	level~\cite{Steiner}).
	\item[(3)]  
	{\bf Monte Carlo methods}: MC is a powerful method for numerical 
	pricing calculation~\cite{Kemna_monte_carlo_asian_opt, Boyle,
	Jackel}. 
	Indeed when the problem dimension becomes high, 
	it is the natural choice respect to binomial or trinomial tree. 
	That because the pricing error, in MC, scales as $1/\sqrt{N}$ 
	(where $N$ is the number of simulations) independently from 
	problem dimension. Moreover, Monte Carlo can be implemented quite 
	straightforward for all path dependent european options, it 
	therefore not surprising that this method is largely used in banks 
	and financial institutions. On the other hand, the convergence 
	speed is low, making MC approach computationally demanding; that 
	have originated a plethora of improved convergence methods and 
	their relatives (reduced variance techniques, quasi Monte Carlo 
	approach based on low discrepancy sequences etc.~\cite{Jackel})
	\item[(4)]  
	{\bf PDE solution}: as well known, the problem of option pricing
	can be formulated in terms of partial differential 
	equation (PDE)~\cite{Black}; therefore a possible strategy, to 
	price derivatives, consists to solve numerically a PDE. A good 
	example of this approach, in the case of path dependent option 
	with discrete sampling, is given in~\cite{Wilmott_pde_1, 
	Wilmott_pde_2, Tavella_pde}. A strong point in favor of this 
	method, is the possibility of considering different processes 
	alternative to geometric brownian motion (GBM). However some 
	care must be taken in the implementation of the method. 
	\item[(5)]  
	{\bf quadrature method}: 
	the option pricing of path dependent european options could be 
	reformulated as a path integral. In the case of discrete fixing 
	dates, this path integral reduces to a multi dimensional integral 
	(whose dimension correspond to the number of observation dates). 
	As shown by some authors~\cite{Aitsahlia_quadrature_barrier_opt, 
	Aitsahlia_quadrature_lookback_opt, Sullivan_quadrature_barrier_opt, 
	Lim_quadrature_asian_opt, Andricopoulos_quadrature_methods,
	Fusai_quadrature_barrier_opt}, often this multiple integral can 
	be further reduced to a series of recursive (nested) one-dimensional 
	integrals (one for each fixing dates), whose solution  leads to 
	an estimation of the option price. \\
	The problem faced by quadrature methods is how to treat accurately 
	the density function involved in each of these one-dimensional 
	nested integrals. Three main different approaches have been 
	proposed in literature:

	\begin{itemize}
		\item[(i)] 
		a basic simple idea~\cite{Aitsahlia_quadrature_barrier_opt, 
		Aitsahlia_quadrature_lookback_opt} consists to discretize, 
		via a grid of points, the density at each time step. Then,
		by adopting a recursive algorithm, the density function at 
		the next time step is numerically calculated basing on the 
		values at the previous step. \\
		This numerical scheme can prove to be expensive, from a 
		computational point of view, as the result accuracy 
		depends directly from the number of points composing
		the grid.
		\item[(ii)] 
		alternative to crude Numerical Recursive Integration (NRI) 
		presented above,
		one can resort to parametric approximations of the real 
		density, by finding the distribution parameters via 
		quadratic fits. For instance, in the case of asian options, 
		a popular choice is to parametrize the density of 
		underlying arithmetic average, at each observation date, 
		as a log-normal distribution. An improvement was suggested 
		by Lim~\cite{Lim_quadrature_asian_opt}, who proposed a more 
		general mixed density approximation in order to take in 
		account the skewness of the distribution which comes up 
		when high value of underlying volatility are considered. \\
		However the results, depend strictly on the arbitrary 
		choice of a particular form used to model real densities. 
		In other word, we do not have a rational criteria in order 
		to individuate the better parametric representation. 
		\item[(iii)] 
		a third alternative consists to combine numerical 
		integration with function approximation. 
		In~\cite{Sullivan_quadrature_barrier_opt}, the authors 
		consider the problem of a barrier option with discrete 
		monitoring dates. They approximate the solution at each 
		observation date by a Chebyshev polynomials and solve the
		integral via Gauss-Legendre quadrature. 
		An alternative approach has been proposed by Fusai et al. 
		in~\cite{Fusai_quadrature_barrier_opt}, again for barrier 
		options. The authors approximate the solution thorough a 
		linear combination of hat functions. All the task consists 
		to relate the coefficients of the linear combination at 
		time $t_{i+1}$ to the coefficients previously computed at 
		time $t_i$. They also provide an error estimate, which 
		depends on the grid spacing used to interpolate the 
		density function. This error is found to be quadratic in 
		spatial discretization.  

	\end{itemize}
\end{itemize}

In this paper we propose a new methodology which belongs to the area of 
quadrature algorithms. The method, we have called perturbative moment 
expansion (PME), focuses the attention on distribution moments instead 
of PDF's. 
The main idea is based on two legs:
\begin{itemize}
	\item[(a)] 
	a generic PDF can be always reconstructed, knowing its 
	moments, by resorting to an extension of Gram-Charlier 
	Series~\cite{Gram-Charlier_series}. More precisely a
	PDF can be expressed as a perturbative series expansion 
	around another (simpler) PDF, by matching all moments (up 
	to a given order) of the original distribution. 
	These arguments are presented in 
	section~\ref{section:PDF_moments_expansion}.

	\item[(b)] 
	on the other hand, basic operations involving PDF's, are simpler 
	in terms of moments. For instance, in 
	section~\ref{section:math_of_PDF}, we show how the convolution 
	product between two distributions, can be reduced to simple 
	arithmetics by reformulating the problem through moments. 

\end{itemize}
Because often the option pricing can be reduced to a recursive 
numerical integrations over density functions, PME permits to 
solve efficiently the problem with little computational efforts 
(section~\ref{section:results}). 
Indeed by including in the calculation just the first four moments 
(mean, volatility, skewness and kurtosis) the option pricing accuracy 
could be already high. \\
In particular in section~\ref{section:results} we show how to implement 
our method for a variety of contingent claims, that is: barrier, asian 
and reverse cliquet options. \\
Finally section~\ref{section:conclusions}, is devoted to discuss briefly 
some conclusions and remarks.

%
\section{Perturbative moments expansion of a probability distribution function} 
\label{section:PDF_moments_expansion}

Among quadrature methods for option pricing, a key problem is to model 
accurately PDF. In this section we will show how it is possible to 
accomplish the task in a very efficient way. \\

Let us consider a generic stochastic variable $x$ with PDF given by $P$. \\
The $P$ moment of order $k$, is defined as:
\begin{equation}
\langle [x-<x>_{P}]^k \rangle_{P} = 
\int_{-\infty}^{+\infty} \, \left (x - <x>_{P} \right)^k \, P(x) \, dx  \;.
\label{eq:def_moment}
\end{equation}
where $<x>$ is the mean of $P$: 
$\langle x \rangle_{P} = \int \, x \, P(x) \, dx $. \\
The normalized moment of order $k$, is defined as follows:
\begin{equation}
\mu^{(k)}_{P} = \frac{\langle [x-<x>_{P}]^k \rangle_{P}}
{\left \{ \langle [x-<x>_{P}]^2 \rangle_{P} \right \}^{\frac{k}{2}}}  \;.
\label{eq:def_normalized_moment}
\end{equation}
By indicating with $\sigma_{P}$ the square root of second moment (i.e. 
the standard deviation of $P$):  
$\sigma_{P} = \left \{ \langle [x-<x>_{P}]^2 \rangle_{P} 
\right \}^{\frac{1}{2}}$, we can always decompose $x$ as: 
\begin{eqnarray}
x &=& <x>_{P} + \sigma_{P} \, y \, , \nonumber \\
P(x) &=& \frac{1}{\sigma_{P}} \, 
\bar P \left (\frac{x-<x>_{P}}{\sigma_{P}} \right ) \, ,
\label{eq:pdf_decomposition}
\end{eqnarray}
where $y$ is a stochastic variable with zero mean and unit variance and  
$\bar P$ denotes its PDF. \\
It turns out that the probability density function $\bar P$, can be 
represented as a series expansion in terms of a polynomial multiplied 
by the normal density $\Phi_{0,1}$ (with unit variance and zero mean). 
The polynomial accounts, in this way, the departure of the original PDF 
from normality. This expansion is known in literature as Gram-Charlier 
Series~\cite{Gram-Charlier_series}. Usually the infinite series is
truncated to a given order, very often up to fourth moment, while higher 
moment corrections are neglected. In such a way it is possible to 
incorporate adjustments in the probability distribution for non-normal 
skewness and kurtosis effects (the former being the third moment and 
accounts for asymmetric tails, while the later corresponds to the fourth 
moment and incorporates, for value higher than three, fatness in the 
tails). \\
In past years some authors~\cite{Jarrow, Corrado_gram, Knight_gram} have 
resorted to Gram-Charlier Series to derive corrections to Black \& Scholes 
formula for plain vanilla options, by including skewness and leptokurticity 
effects in the distribution of stock returns. \\
The Gram-Charlier Series, truncated to the fourth moment, reads:
\begin{equation}
\bar P(x) = \left [ 1 + \frac{\mu_3}{6} \, H_3(x) + 
\frac{\mu_4 -3}{24} \, H_4(x) + ..... \right ] \, \Phi_{0,1}(x) \; ,
\label{eq:gram_charlier_series_first_4_moments}
\end{equation}
where $\Phi_{0,1}$ is the normal distribution with zero mean and unit 
variance, $\mu_3$ and $\mu_4$ are respectively the skewness and kurtosis 
of $\bar P$ and $\big\{ H_n(x) \big\}_{n \in \cal N}$ denote the Hermite 
polynomials~\cite{Gradshteyn}. \\
More generally, we can write:
\begin{equation}
\bar P(x) \approx \left ( \sum_{i=0}^{K} c_i \, x^i \, \right ) \, 
\Phi_{0,1}(x) \; ,
\label{eq:gram_charlier_series}
\end{equation}
where the coefficients $c_i$ (up to order $K$) can be easily computed by 
imposing the equivalence of the first $K+1$ moments of both sides of 
equation~(\ref{eq:gram_charlier_series}): 
\begin{equation}
\sum_{i=0}^{K} c_i \, \langle x^{i+j} \rangle_{\Phi_{0,1}} = 
\langle x^j \rangle_{\bar P}  \; \; \; , \mbox{for: } 
\; j=0, 1, 2, ..... K \, ,
\label{eq:linear_system_eq_for_moments}
\end{equation}
where we have indicated with $<x^i>_{\Phi_{0,1}}$ the gaussian moment 
of order $i$. Remember that for a gaussian distribution with zero mean 
and unit variance: 
\begin{eqnarray}
<x^j>_{\Phi_{0,1}} & = & 0 \qquad \qquad \qquad \qquad \qquad 
\mbox{if $j$ is odd} \, , \nonumber \\
<x^j>_{\Phi_{0,1}} & = & 1 \cdot 3 \cdot 5 \cdot ....... \cdot (j-1) 
\quad \; \mbox{if $j$ is even} \, ,
\label{eq:gaussian_moments}
\end{eqnarray}
moreover, $\bar P$ is a PDF with zero mean and unit variance, therefore: \\
$ <x^0>_{\bar P}  =  1 $, $ <x^1>_{\bar P}  =  0 $ and 
$<x^2>_{\bar P}  =  1 $. \\
By reverting the linear system 
equations~(\ref{eq:linear_system_eq_for_moments}), the coefficients 
$\{c_i\}_{0}^{K}$ can be easily found in terms of a linear combination 
of the first $K+1$ moments of $\bar P$. 
Therefore, in some sense, the equation~(\ref{eq:gram_charlier_series}), 
can be regarded as a perturbative moment expansion (PME) of $\bar P$, 
around gaussian function $\Phi_{0,1}$. The convergence of the series
to $\bar P$ is guaranteed by some theorems~\cite{Cramer}. \\
It is straightforward to extend the equations~(\ref{eq:gram_charlier_series})
and~(\ref{eq:linear_system_eq_for_moments}), to the case where 
$\Phi_{0,1}$, is substituted with a generic PDF, $\varphi$ (again with 
zero mean and unit variance). Of course, it is intuitively, that the 
quality of approximation in~(\ref{eq:gram_charlier_series}), at a given 
order $K$, depends on the ``distance'' between $\bar P$ and $\varphi$. 
%
%

%
\section{The convolution of probability distributions via moments 
calculation} \label{section:math_of_PDF}

Let us consider two independent stochastic variables $x_1$ and $x_2$, 
each of theme characterized by a different probability distribution 
function $P_1$ and $P_2$. The sum of the two variables: $z=x_1+x_2$ is 
again a stochastic variable with a probability distribution, $P$, given 
by the convolution of $P_1$ and $P_2$:
\begin{equation}
P(z) = \int \int_{x_1+x_2=z} \, P_1(x_1) \, P_2(x_2) \, dx_1 \, dx_2 = 
\int_{-\infty}^{+\infty} P_1(x_1) \, P_2(z-x_1) \, dx_1 \,.
\label{eq:pdf_convolution}
\end{equation}
Starting from the above equation, it easy to find the composition law 
connecting the $P$ moments to $P_1$ and $P_2$ moments:
\begin{eqnarray}
&& \langle z \rangle_P = \langle x_1 \rangle_{P_1} 
+ \langle x_2 \rangle_{P_2} \, \nonumber \\
&& \nonumber \\
&& \langle \left[z - <z> \right]^n \rangle_P \, =  
\label{eq:moment_composition_law} \\
&& = \sum_{\nu=0}^{n} \, {n \choose \nu } 
\, \langle \left[ x_1 - <x_1>_{P_1} \right] ^{\nu} \rangle_{P_1} 
\cdot \langle \left[ x_2 -<x_2>_{P_2} \right]^{n-\nu} \rangle_{P_2} 
\,, \nonumber
\end{eqnarray}
On the opposite of eq.~(\ref{eq:pdf_convolution}), the moments composition 
law is quite simple and straightforward. Therefore an alternative approach 
to compute the convolution product between $P_1$ and $P_2$, in order to 
find out $P$, consists to compute $P$ moments (up to a certain order) by 
using~(\ref{eq:moment_composition_law}) and then reverting to $P$ by 
considering the perturbative expansion~(\ref{eq:gram_charlier_series}). 
The potentiality of this technique could be appreciated by considering the 
following problem: given a stochastic variable $z$ characterized by a 
probability distribution $P$, we ask to find out the PDF, $P$, such that 
the convolution: $P_{1/2} \circ P_{1/2}$ is equal to $P$ (in other word we 
are ask to extract the square root of $P$ respect to convolution product). 
This problem could be hard from a numerical point of view by considering 
the eq.~(\ref{eq:pdf_convolution}) but becomes quite simple by resorting 
to PME and moments composition law. Indeed, within PME scheme, the problem
reduces to simple algebra: once we have got the moments characterizing the 
square root of $P$, via inversion of eq.~(\ref{eq:moment_composition_law}), 
it is simple to reconstruct the corresponding PDF by using 
equation~(\ref{eq:gram_charlier_series}).
%

If we consider the sum of $n$ i.i.d. variables, with distribution function 
$P_1$, the PDF of the sum, is given by $P_n= \prod_{1}^{n} P_1$ (where 
$\prod$ refers to the convolution product). In terms of moments 
this equation becomes:
\begin{eqnarray}
&& \langle \left [ x-<x>_{P_n} \right]^k \rangle_{P_n} =
n \, \langle \left [ x-<x>_{P_1} \right]^k \rangle_{P_1} + \nonumber \\
&& + \sum_{\nu=1}^{k-1} \, {k \choose \nu} \,
\left \{
\sum_{j=1}^{n-1} \, \langle \left [ x-<x>_{P_j} \right]^{k-\nu} \rangle_{P_j} 
\right \} \, \langle \left [ x-<x>_{P_1} \right]^{\nu} \rangle_{P_1} \, .
\label{eq:moment_composition_law_for_iid}
\end{eqnarray}
Equation~(\ref{eq:moment_composition_law_for_iid}), permits to evaluate 
iteratively, all $P_n$ moments. More precisely, a solution can be easily 
found for $k=2$, then, starting from the knowledge of $k-1$-moment of 
$P_n$, we have a rule to construct the next $k$-moment of $P_n$. \\  
As an example, we can solve iteratively 
equation~(\ref{eq:moment_composition_law_for_iid}), for first values 
of $k$ (i.e. $k=2$, $k=3$ and $k=4$, which give the evolution laws, 
respectively, for volatility, skewness and kurtosis): 
\begin{eqnarray}
\langle \left [ x-<x>_{P_n} \right]^2 \rangle_{P_n} &=& n \, 
\langle \left [ x-<x>_{P_1} \right]^2 \rangle_{P_1} \, ,
\label{eq:moment_composition_law_vol} \\
\mu^{(3)}_{P_n} &=& \frac{ \mu^{(3)}_{P_1} }{n^{1/2}} \, ,
\label{eq:moment_composition_law_skewness} 
\\
\mu^{(4)}_{P_n} &=& 3+ \frac{\mu^{(4)}_{P_1}-3}{n} \, .
\label{eq:moment_composition_law_kurtosis} 
\end{eqnarray}
More generally, starting from eq.~(\ref{eq:moment_composition_law_for_iid}), 
it possible to derive a perturbative series expansion 
of reduced moments $\mu^{(k)}_{P_n}$, in powers of 
$\frac{1}{\sqrt{n}}$. 
In equation~(\ref{eq:moment_expansion_1_over_n}), we show the 
first corrections beyond gaussian result ($n=\infty$): 
\begin{eqnarray}
\mu^{(k)}_{P_n} &=& \mu^{(k+1)}_{\Phi_{0,1}} \left [ \frac{k-1}{3!} \, 
\frac{\mu^{(3)}_{P_1}}{n^{1/2}} 
+ O \left( \frac{1}{n^{3/2}} \right )
\right ] \qquad \qquad \quad  \mbox{k odd} \, , \nonumber \\
&& 
\label{eq:moment_expansion_1_over_n} \\
\mu^{(k)}_{P_n} &=& \mu^{(k)}_{\Phi_{0,1}} \left [ 1 + 
\frac{k \, (k-2)}{4!} 
\, \frac{\mu^{(4)}_{P_1}-3}{n} + 
O \left( \frac{1}{n^{2}} \right ) \right ] \quad  \; \; \mbox{k even} 
\, , \nonumber 
\end{eqnarray}
where $\mu^{(k)}_{\Phi}$ represents the normalized gaussian moment of 
order $k$. \\
Formula~(\ref{eq:moment_expansion_1_over_n}), shows that for large $n$, 
the PDF, $P_n$, converges (according to central limit theorem) to a 
gaussian distribution and the first two corrections to the asymptotic 
limit, depends only from $\mu^{(3)}_{P_1}$ and $\mu^{(4)}_{P_1}-3$ 
(this result is the PME version of a famous theorem obtained by 
Kolmogorov and Gnedenko, for probability distributions, in 
1954~\cite{Kolmogorov}\footnote{The theorem provide an asymptotic 
series expansions of distribution function $P_n$ as:
\begin{equation}
P_n(x) - \Phi_{0,1}(x) = \Phi_{0,1}(x) \, \sum_{j=0}^{+\infty} 
\, \frac{Q_j(x)}{n^{\frac{j}{2}}}
\label{eq:pdf_expansion_1_over_n}
\end{equation}
where $Q_j$ is a polynomial whose coefficients depend uniquely from the
first $j+2$ moments of $P_n$. The explicit form of $\{ Q_j \}$ can be 
found in~\cite{Kolmogorov}}).

%
%
\section{Perturbative moment expansion for pricing European style
options} 
\label{section:results}

The option pricing of path dependent european options with discrete 
time monitoring, is equivalent to solve a multiple integral. As 
shown in literature~\cite{Aitsahlia_quadrature_barrier_opt, 
Aitsahlia_quadrature_lookback_opt, Andricopoulos_quadrature_methods, 
Sullivan_quadrature_barrier_opt, Lim_quadrature_asian_opt, 
Fusai_quadrature_barrier_opt}, often this multiple integral can be reduced
to nested one-dimensional integrals (one for each fixing dates). 
The problem faced by quadrature methods is how to treat accurately the 
density functions involved in such integrals (see the introduction). \\
In this paper, we propose a new scheme for modeling density functions, 
avoiding arbitrary choices of parametric functions as well as large grids 
which can lead easily to high computational costs. \\ 
The basic idea relies from one hand on perturbative moment expansion of a 
generic PDF (equation~(\ref{eq:gram_charlier_series})) and from the other hand 
on moments composition laws derived in section~\ref{section:math_of_PDF}. \\
We expose PME technique in practice, by considering three different kind of 
path dependent options:
\begin{itemize}
	\item[(i)] 
	asian options with discrete fixings 
	(chapter~\ref{subsection:asian_option_pricing});

	\item[(ii)] 
	reverse cliquet options
	(chapter~\ref{subsection:reverse_cliquet_option_pricing});

	\item[(iii)] 
	barrier options with discrete monitoring dates
	(chapter~\ref{subsection:barrier_option_pricing});

\end{itemize}
for each of them, we present results obtained with PME approach. A 
comparison with other popular techniques (Monte Carlo, Quasi Monte Carlo, 
recursive numerical integration, binomial tree etc.) is also reported.

%
\subsection{Asian options with discrete fixings} 
\label{subsection:asian_option_pricing}

\subsubsection{Asian options: contract description} 
\label{subsubsection:asian_contract_des}

We consider a call asian option written on a stock with initial value 
$S_0$ and volatility $\sigma$. The interest rate, $r$, is assumed, for 
simplicity, to be a constant (indeed in our scheme this hypothesis is 
not necessary). \\
The contract pay-off is defined as:
\begin{equation}
\mbox{Pay-off}_{\mbox{{\tiny Asian}}} = 
\mbox{Max} \left( \frac{\sum_{i=0}^{m} S_i}{m+1} \,  - E, 0 \right) \; ,
\label{eq:pay_off_asian_opt}
\end{equation}
where $S_i$ is the stock price at time $t_i = i \; T/m $, $m$ is the 
number of equally spaced intervals and T indicates the time to 
maturity. \\

\subsubsection{Asian options: PME problem formulation} 
\label{subsubsection:asian_pme_formulation}

Equity price at time $t_j$, can be written as:
\begin{equation}
S_j = S_0 \, e^{\, \sum_{i=1}^{j} \left( \rho_i + v_i \, e_i \right) } \; ,
\label{eq:S_i_asian_opt}
\end{equation}
where: $e_i$'s are independent $N(0,1)$ random variables with zero mean 
and unit variance; $\rho_i$ and $v_i^2$ are respectively the mean and 
variance of stock log-returns calculated between $t_{i-1}$ and $t_i$:
\begin{eqnarray}
\rho_i &=& \left( r - \frac{\sigma^2}{2} \right ) \, 
\left ( t_i-t_{i-1} \right )  \, , \nonumber \\
& & 
\label{eq:mean_sigma_asian_opt} \\
v_i &=& \sigma \, \sqrt{ t_i-t_{i-1} } \, . \nonumber
\end{eqnarray}
Now, we introduce the stochastic variables, $a_i$, defined recursively 
as follows:
\begin{eqnarray}
a_0 &=& \log{ \left( 1 + e^{a_{1}} \right)} \; , 
\label{eq:a_0_definition} \\
a_i &=& \rho_i + v_i \, e_i + \log{ \left( 1 + e^{a_{i+1}} \right)} \; , 
\label{eq:a_i_definition} \\
a_{m} &=& \rho_m + v_m \, e_m  \; . 
\label{eq:a_m_definition}
\end{eqnarray}
$a_i$'s can be also re-written as 
\begin{equation}
a_i = \hat \rho_{i} + \hat v_{i} \, \hat e_i \; ,
\label{eq:a_i_decomposition}
\end{equation}
where $\hat \rho_{i}$ and $\hat v_{i}^2$ are the mean and variance of $a_i$ 
and $\{ \hat e_i \}$ are new stochastic variables with zero mean and unit 
variance. $\hat e_i$'s variables, on the opposite of $e_i$'s, are strictly 
related one to each other; more important, their PDF's, (referred in the 
following as $P_{\hat e_i}$), are in principle not normal. \\
The option value, $c_{\mbox{{\tiny asian}}}$, can be then calculated as:
\begin{equation}
c_{\mbox{{\tiny asian}}} = e^{-r \, T} \, \int_{-\infty}^{+\infty} \mbox{Max} 
\left [ \frac{e^{\hat \rho_{0} + \hat v_{0} \, x}}{m+1} - E, 0 \right ] 
\, P_{\hat e_0}(x) \, dx \, .
\label{eq:asian_opt_pricing}
\end{equation}
Therefore, the pricing of an asian option, with discrete fixings, is 
reduced to the calculation of a one-dimensional integral, with a PDF 
given by $P_{\hat e_0}$. The problem is therefore reduced to compute 
recursively: $\hat \rho_{i}$, $\hat v_{i}$ and $P_{\hat e_i}$ backward,
until $i=0$ is reached. 
This task could be well accomplished, by using a perturbative moment 
expansion (PME) of $P_{\hat e_i}$ around normal distribution $\Phi_{0,1}$. 
In detail, the iterative PME scheme for an asian option is defined as 
follows:
\begin{itemize}
	\item[(A)] 
	fix the number of moments, $l$, to be used in PME algorithm;

	\item[(B)] 
	let us start from $i=m$, Eqs.~(\ref{eq:a_m_definition}) 
	and~(\ref{eq:a_i_decomposition}) give 
	$\hat \rho_{m} = \rho_m$, $\hat v_{m} = v_m$ and 
	$P_{\hat e_m}=\Phi_{0,1}$;

	\item[(C)] 
	proceeding backward, for $i<m$, let us introduce a dummy 
	stochastic variable:
	$y_{i+1} = \log \left( 1 + e^{\hat \rho_{i+1} + \, \hat v_{i+1} \, 
	\hat e_{i+1}} \right)$. Knowing $\hat \rho_{i+1}$, $\hat v_{i+1}$ 
	and $P_{\hat e_{i+1}}$, we can compute, numerically, all moments 
	(up to order $l$) of $y_{i+1}$. \\
	Observing that $a_i$ is the sum of two independent stochastic
	variables (i.e. $\rho_i + v_i \, e_i$ and $y_{i+1}$), we can 
	calculate, by applying the moments composition rule derived in 
	eq.~(\ref{eq:moment_composition_law}): $\hat \rho_{i}$, 
	$\hat v_{i}$ and all moments (up to $l$) of $P_{\hat e_i}$. \\
	Finally, by resorting to eq.~(\ref{eq:gram_charlier_series}), 
	we are able to reconstruct $P_{\hat e_i}$ from its moments.

	\item[(D)] 
	when $i=0$, is reached, we can compute the option value via 
	eq.~(\ref{eq:asian_opt_pricing}).
\end{itemize}
The higher the number of moments we include in the calculation, the 
higher will be the precision of option value estimate.
However, with just including kurtosis and skewness (and neglecting higher 
moments corrections), we have got excellent results (the percentage error 
is less than $0.1$ \%, see tables~\ref{Tab:asian_PME_versus_other_1} 
and~\ref{Tab:asian_PME_versus_other_2}). \\

\subsubsection{Asian options: PME option pricing, a comparison with 
other techniques} 
\label{subsubsection:asian_numerical_results}

%
%
%
%
%
{
 
\begin{table}[h]
\tiny

\begin{center}

\begin{tabular}{|l|l|l|l|l|}
\hline
\multicolumn{5}{|c|} {} \\
\multicolumn{5}{|c|}{\bf \large Asian option pricing with different techniques.} \\
\multicolumn{5}{|c|} {} \\
\hline \hline
                        &                         &                         &                         &                     	\\
                        & $\sigma = 5$ \%         &  $\sigma = 10$ \%       &  $\sigma = 30$ \%       &  $\sigma = 50$ \%   	\\
                        &                         &                         &                         &                     	\\
\hline \hline
                        &                         &                         &                         &                     	\\
PME ($l=4$)             &  $4.30799$              &  $4.90899$              &  $8.79859$              &  $12.96664$         	\\
                        &                         &                         &                         &                     	\\
\hline
                        &                         &                         &                         &                     	\\
PME ($l=6$)             &  $4.30798$              &  $4.90899$              &  $8.80149$              &  $12.97995$         	\\
                        &                         &                         &                         &                     	\\
\hline
                        &                         &                         &                         &                     	\\
PME ($l=8$)             &  $4.30798$              &  $4.90899$              &  $8.80142$              &  $12.98102$         	\\
                        &                         &                         &                         &                     	\\
\hline
                        &                         &                         &                         &                     	\\
PME ($l=10$)            &  $4.30798$              &  $4.90899$              &  $8.80153$              &  $12.98124$         	\\
                        &                         &                         &                         &                     	\\
\hline
                        &                         &                         &                         &                     	\\
PME ($l=20$)            &  $4.30798$              &  $4.90899$              &  $8.80151$              &  $12.98097$         	\\
                        &                         &                         &                         &                     	\\
\hline
                        &                         &                         &                         &                     	\\
MC                      & $4.30795$ +/- $0.00003$ &  $4.9089$ +/- $0.00014$ & $8.80095$ +/- $0.00062$ &  $12.980$ +/- $0.0012$  \\
                        &                         &                         &                         &                     	\\
\hline
                        &                         &                         &                         &                     	\\
RNI                     &  $4.308$                &  $4.909$                &  $8.801$                &  $12.980$           	\\
                        &                         &                         &                         &                     	\\
\hline 
                        &                         &                         &                         &                     	\\
MDA                     &  $4.309$                &  $4.911$                &  $8.811$                &  $12.979$           	\\
                        &                         &                         &                         &                     	\\
\hline
\end{tabular}
\end{center}
\caption{
	{\small
	Option pricing for an asian option with: 
	$S_0=100$, $r=9$ \%, $T=1$ year, $E=100$ and $m=52$ (number 
	of intervals). 
	The table shows option values computed with different 
	techniques: 
	PME - Perturbative Moment Expansion (with different values of $l$), 
	MC - Monte Carlo ($10^8$ scenarios with antithetic variable technique),
	RNI - Recursive Numerical Integration (results are due to Lim~\cite{Lim_quadrature_asian_opt}) and
	MDA - Mixed Density Approximation (results are due to Lim~\cite{Lim_quadrature_asian_opt})
	} 
}
\label{Tab:asian_PME_versus_other_1}
\end{table}
}
%
%
In order to compare our results with literature, we have considered 
two set of parameter values, reported in the captions of 
table~\ref{Tab:asian_PME_versus_other_1} 
and~\ref{Tab:asian_PME_versus_other_2}. \\
In table~\ref{Tab:asian_PME_versus_other_1} we compare PME results 
(number of moments, $l$, ranging from $4$ up to $20$) against other 
popular techniques: 
(i) Monte Carlo (MC) with $N=10^8$ scenarios, where we have resort 
to antithetic variables technique in order to reduce variance errors; 
(ii) Recursive Numerical Integration (RNI) and 
(iii) Mixed Density Approximation (MDA), which represents a modification 
of traditional RNI algorithm in which an appropriately parametrization 
of distribution functions is used. \\
%
%
%
{
 
\begin{table}[h]
\tiny

\begin{center}

\begin{tabular}{|l|l|l|l|}
\hline
\multicolumn{4}{|c|} {} \\
\multicolumn{4}{|c|}{\bf \large Asian options pricing with different techniques.} \\
\multicolumn{4}{|c|} {} \\
\hline \hline
                        &                           &                          &                     	  \\
                        & $\sigma = 10$ \%          &  $\sigma = 20$ \%        &  $\sigma = 40$ \%   	  \\
                        &                           &                          &                     	  \\
\hline \hline
                        &                           &                          &                     	  \\
PME ($l=4$)             &  $1.845289$               &  $2.920988$              &  $5.14583$               \\
                        &                           &                          &                     	  \\
\hline
                        &                           &                          &                     	  \\
PME ($l=6$)             &  $1.845298$               &  $2.921097$              &  $5.14677$               \\
                        &                           &                          &                     	  \\
\hline
                        &                           &                          &                     	  \\
PME ($l=8$)             &  $1.845297$               &  $2.921095$              &  $5.14677$               \\
                        &                           &                          &                     	  \\
\hline
                        &                           &                          &                     	  \\
PME ($l=10$)            &  $1.845298$               &  $2.921097$              &  $5.14678$               \\
                        &                           &                          &                     	  \\
\hline
                        &                           &                          &                       	  \\
PME ($l=20$)            &  $1.845298$               &  $2.921097$              &  $5.14678$               \\
                        &                           &                          &                     	  \\
\hline
                        &                           &                          &                     	  \\
MC                      &  $1.84517$ +/- $0.00008$  &  $2.9209$ +/- $0.00018$  &  $5.1463$ +/- $0.00040$  \\
                        &                           &                          &                     	  \\
\hline
                        &                           &                          &                     	  \\
RNI                     &  $ 1.845 $                &  $ 2.921 $               &  $ 5.146 $          	  \\
                        &                           &                          &                     	  \\
\hline 
                        &                           &                          &                     	  \\
FSG                     &  $ 1.869 $                &  $ 2.960 $               &  $ 5.218 $          	  \\
                        &                           &                          &                     	  \\
\hline
\end{tabular}
\end{center}
\caption{
	{\small
	Option pricing for an asian option with: 
	$S_0=100$, $r=10$ \%, $T=\frac{91}{365}$ year, $E=100$ and $m=91$. 
	The table shows option value computed with different approaches: 
	PME - Perturbative Moment Expansion (with different values of $l$), 
	MC - Monte Carlo ($10^8$ simulations with antithetic variable technique),
	RNI - Recursive Numerical Integration (results are due to Lim~\cite{Lim_quadrature_asian_opt}),
	FSG - Forward Shooting Grid (Barraquand and Pudet in~\cite{Barraquand_tree}).} 
	}
\label{Tab:asian_PME_versus_other_2}
\end{table}
}
%
%
%
%
In table~\ref{Tab:asian_PME_versus_other_2}, a comparison of PME with MC, 
RNI and  binomial tree (adopting forward shooting grid algorithm) of 
Barraquand and Pudet~\cite{Barraquand_tree} is also presented. \\
%
%
%
%
%
\begin{figure}[hbtp]
\begin{center}
\epsfig{file=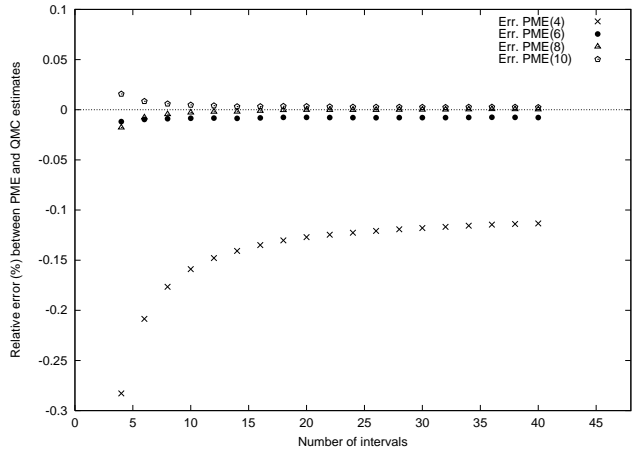,scale=0.34,angle=-90}
\caption{
  { \small
  Percentage error between QMC result ($N=2^{27}-1$) and PME results 
  ($l=4 \div 10$) for an asian option with: 
  $S_0=100$, $r=9$ \%, $T=1$ year, $E=100$, $\sigma = 50$ \% and $m$ 
  (fixing frequency) ranging from $4$ up to $40$.
  }
}
\label{Fig:asian_PME_versus_QMC}
\end{center}
\end{figure}
%
%
%
%
Finally, in figure~\ref{Fig:asian_PME_versus_QMC}, we show the relative 
error (in percent) between PME and Quasi Monte Carlo (QMC) 
results\footnote{In spite of the fact that QMC method does not permit to 
estimate the error, we have preferred this algorithm to simple MC because 
a lower number of scenarios are required to obtain an estimate of a given 
precision.}. 
QMC estimates are based on $N=2^{27}-1$ simulation series, making use of 
Sobol low-discrepancy sequences. \\
The results obtained, show that PME method give excellent results (error 
being less than 1\%) also in the case we have retained few moments 
($l=4$, i.e. just kurtosis and skewness). \\
The main advantage of PME respect to other numerical techniques are:

\begin{itemize}

	\item[(i)] 
	PME method, as all quadrature techniques, does not 
	suffer from increasing fixing observations number or volatility  
	values where, on the opposite, the precision of MC estimates 
	deteriorates when $\sigma$ or $m$ increase.

	\item[(ii)] 
	Once the final PDF is obtained, option prices can be easily 
	computed for any $S_0$ and strike price $E$, moreover that gives 
	the opportunity to compute immediately the delta greek of the  
	option. 

	\item[(iii)] 
	Respect to the methodology used in~\cite{Lim_quadrature_asian_opt}, 
	PME does not require to guess or make any hypothesis about the 
	PDF form, nor to compute PDF on a grid of points. 
	Indeed we have only to compute the first moments (e.g. for $l=4$: 
	mean, variance, kurtosis and skew, i.e. just four numbers) at 
	each fixing date, making the option evaluation quite fast. 
	For a general discussion about the advantage of PME algorithm, we 
	remand to section~\ref{section:conclusions}.

\end{itemize}

%
\subsection{Reverse cliquet options} 
\label{subsection:reverse_cliquet_option_pricing}

\subsubsection{Reverse cliquet options: contract description} 
\label{subsubsection:reverse_cliquet_contract_des}

A reverse cliquet option is a product that typically have a maximum 
and minimum pay-out, which depends on the sum of negative performances 
of the underlying asset. \\
A typical pay-out might be written as:
\begin{equation}
\mbox{Pay-off}_{\mbox{{\tiny Rev. cliquet}}} = 
\mbox{Max} \left [ L, H+ \sum_{i=1}^{m} \mbox{Min} \left ( 
\frac{S_i-S_{i-1}}{S_{i-1}}, 0 \right ) \right ] \; ,
\label{eq:pay_off_reverse_cliquet_opt}
\end{equation}
where the cliquet has $m$ periods, $S_i$ is the underlying price at the 
end of the i'th period $(t_{i-1}, t_i)$, the strike level for the first 
period being S0 (i.e. the spot price). $L$ is the global floor (usually 
set to zero) and corresponds to the minimum amount the investor will 
receive at the expiration date. $H$ is the maximum coupon payable by the 
option. \\
The risk profile embedded in this option is clear: the investor is long 
the skew (if the skewness of stock price returns increases, it is less
probable to have negative performances and therefore the contract value 
increases) and short volatility (indeed if volatility grows the contract 
price declines). As we will see, the PME approach permits to calculate 
at each step the effective volatility and skewness of the sum of the 
underlying negative performances (also in presence of non log-normal 
process), making more clear the risk profile.

\subsubsection{Reverse cliquet options: PME problem formulation} 
\label{subsubsection:reverse_cliquet_pme_formulation}

The problem formulation in terms of perturbative moment expansions, 
becomes in this particular case quite simple. \\
Let us introduce the stochastic variables, 
$R_i=\mbox{Min} \left (\frac{S_i-S_{i-1}}{S_{i-1}}, 0 \right )$,
i.e. the minimum value between stock performance and zero. 
If we indicate with $P_i$ the PDF of variables $R_i$, the density
function
of the sum of all negative performances ($\sum_{i=1}^{m} R_i$), is given 
by:
\begin{equation}
P = \prod_{i=1}^{m} P_i  \; ,
\label{eq:reverse_cliquet_conv_product}
\end{equation}
where the symbol, $\prod$, refers to convolution product. \\
The equation~(\ref{eq:reverse_cliquet_conv_product}), which is indeed a 
multiple integral, can be simplified, by reformulating the problem in 
terms of PDF's moments (see section~\ref{section:math_of_PDF}). 
More precisely, the $P$ moments can be computed by combining, iteratively, 
the $P_i$ moments via equation~(\ref{eq:moment_composition_law}). 
On the other hand, the $P_i$ moments can be easily calculated as 
follows:
\begin{eqnarray}
&& R_i(x) = \mbox{Min} \, \left[ e^{-\left ( r - \frac{\sigma^2}{2} \right ) 
\, \delta t_i + \sigma \, \sqrt{\delta t_i} \, x} -1, 0 \right ] \, , 
\quad x \sim N(0,1)	\nonumber		\\
&& \langle R_i \rangle_{P_i}  =  \int \, R_i(x) \, \Phi_{0,1}(x) 
\, dx \, ,  
\label{eq:pdf_moment_negative_performance} \\
&& \langle \left[R_i - <R_i>_{P_i} \right]^k \rangle_{P_i}   =  
\int \, \left [ R_i(x) - <R_i>_{P_i} \right ]^k \, \Phi_{0,1}(x) 
\, dx \, . \nonumber
\end{eqnarray}
Observe that the above integrals, can be easily ``solved'' in terms of 
a sum of cumulative normal functions. \\
In equation~(\ref{eq:pdf_moment_negative_performance}), $\Phi_{0,1}(x)$ 
represents the normal density distribution with zero mean and unit 
variance and $\delta t_i = t_i-t_{i-1} $. \\
Once the $P$ moments have been calculated, by resorting to 
eq.~(\ref{eq:gram_charlier_series}) , it is possible to reconstruct the
corresponding probability distribution function as a series expansion 
around normal density, $\Phi_{0,1}$.
Then, the option price can be easily calculated as:
\begin{equation}
\mbox{option price}_{\mbox{{\tiny Rev. Cl.}}} = 
e^{-r \, T } \, \mbox{Max} \, 
\left[ L, \int  \, (H+x) \, P(x) \, dx \right ] \; .
\label{eq:reverse_cliquet_opt_pricing_with_PME}
\end{equation}

Note that:
\begin{itemize}
	\item[(i)] 
	if time intervals are constant the variables $R_i$ are i.i.d. and 
	therefore their moments can be computed once, making the option 
	price evaluation quite fast (the laws governing moments evolution 
	for i.i.d. permits to compute $P$ moments by simple arithmetics, 
	see equations~(\ref{eq:moment_composition_law_for_iid}), 
	(\ref{eq:moment_composition_law_skewness}) 
	and~(\ref{eq:moment_composition_law_kurtosis}).

	\item[(ii)] 
	the inclusion of kurtosis and skewness effects in the process 
	governing stock price variations does not require any modification 
	to the algorithm, indeed it would be sufficient to substitute in 
	eq.~(\ref{eq:pdf_moment_negative_performance}) to $\Phi_{0,1}$
	the PDF of a non log-normal process.

	\item[(iii)] 
	by increasing the number of intervals, according to 
	equation~(\ref{eq:moment_expansion_1_over_n}), the PDF describing 
	the sum of negative performances, converges to a gaussian 
	distribution (see eq.~(\ref{eq:moment_expansion_1_over_n})). 
	We can expect therefore, that the Gram-Charlier series truncation, 
	used in eq.~(\ref{eq:reverse_cliquet_opt_pricing_with_PME}) to
	model $P(x)$, will become asymptotically exact for large value 
	of $m$.  

\end{itemize}

\subsubsection{Reverse cliquet options: PME option pricing, a comparison 
with other techniques} 
\label{subsubsection:reverse_cliquet_numerical_results}

In table~\ref{Tab:reverse_cliquet_PME_versus_QMC}, we present PME results 
compared with QMC for different numbers of equally spaced intervals, $m$, 
(ranging from $4$ up to $36$). The time intervals are maintained fixed to 
$1/12$ year. In order to make results comparable, we have considered 
$H = m \, h$ ($h$ constant) and option prices (in \%) rescaled by a 
factor $1/m$. 
%
%
%
{
 
\begin{table}[h]
\tiny

\begin{center}

\begin{tabular}{|l|l|l|l|l|}
\hline
\multicolumn{5}{|c|} {} \\
\multicolumn{5}{|c|}{\bf \large Reverse cliquet options pricing } \\
\multicolumn{5}{|c|} {\bf \large with different techniques.}\\
\multicolumn{5}{|c|} {} \\
\hline \hline
                        &                     &                     &                     &                     \\
                        & $m = 4 $            &  $m = 12$           &  $m = 24$           &  $m = 36$           \\
                        &                     &                     &                     &                     \\
\hline \hline
                        &                     &                     &                     &                     \\
PME ($l=4$)             &  $1.41681$  \%      &  $1.01701$  \%      &  $0.82935$  \%      &  $0.72502$   \%     \\
                        &                     &                     &                     &                     \\
\hline
                        &                     &                     &                     &                     \\
PME ($l=6$)             &  $1.42984$  \%      &  $1.01755$  \%      &  $0.82890$  \%      &  $0.72465$   \%     \\
                        &                     &                     &                     &                     \\
\hline
                        &                     &                     &                     &                     \\
PME ($l=8$)             &  $1.43915$  \%      &  $1.01896$  \%      &  $0.82914$  \%      &  $0.72470$   \%     \\
                        &                     &                     &                     &                     \\
\hline
                        &                     &                     &                     &                     \\
PME ($l=10$)            &  $1.43730$  \%      &  $1.01769$  \%      &  $0.82891$  \%      &  $0.72467$   \%     \\
                        &                     &                     &                     &                     \\
\hline
                        &                     &                     &                     &                     \\
PME ($l=20$)            &  $1.43417$  \%      &  $1.01798$  \%      &  $0.82898$  \%      &  $0.72469$   \%     \\
                        &                     &                     &                     &                     \\
\hline
                        &                     &                     &                     &                     \\
QMC                     &  $1.4336$   \%      &  $1.0179$   \%      &  $0.82898$  \%      &  $0.72469$   \%     \\
                        &                     &                     &                     &                     \\
\hline
\end{tabular}
\end{center}
\caption{
	{\small
	Option pricing for a reverse cliquet option with: 
	$S_0=100$, $r=9$ \%, $\sigma = 30$ \%, 
	$\delta_t=t_i-t_{i-1}=1/12$ year, $L=0$ and 
	$m=4$, $12$, $24$, $36$. 
	In order to make results comparable, we have considered 
	$H = m \, h$ ($h=4$ \%) and option prices (expressed in \%) 
	have been rescaled by a factor $1/m$.
	The table shows option values computed with different techniques: 
	PME - Perturbative Moment Expansion (with different values of $l$) and 
	QMC - Quasi Monte Carlo ($2^{27}-1$ simulations). 
	}
}
\label{Tab:reverse_cliquet_PME_versus_QMC}
\end{table}
}
%
%
PME results reported in table~\ref{Tab:reverse_cliquet_PME_versus_QMC}, 
show an excellent agreement with QMC estimates, at least for $m \ge 12$, 
even retaining only few moments in the perturbative expansion (the error 
is less than $0.1$ \% with just the first four moments). \\

%
%
%
\begin{figure}[hbtp]
\begin{center}
\epsfig{file=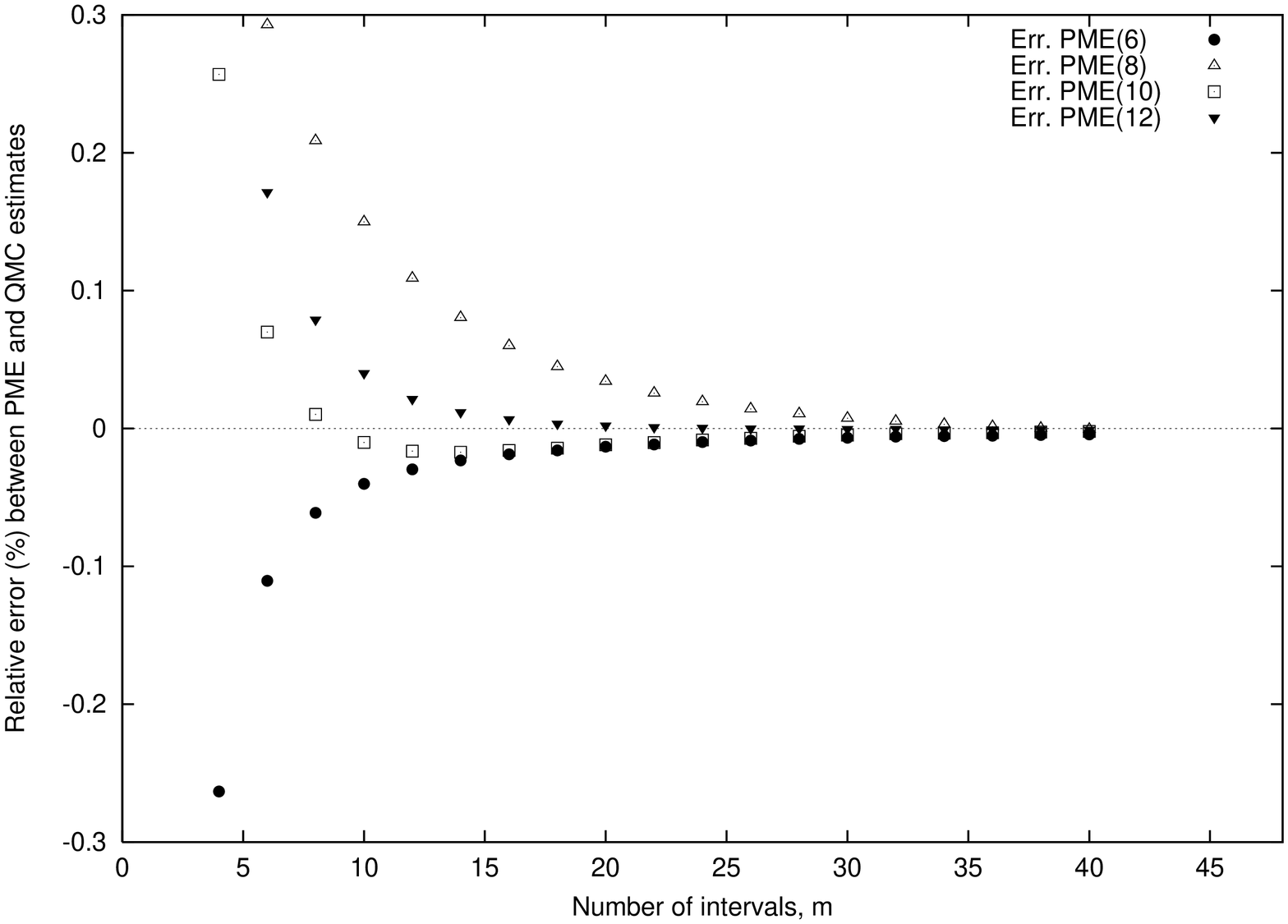,scale=0.34,angle=-90}
\caption{
  { \small
  Reverse cliquet option: the graph shows the relative error between 
  PME and QMC estimates, versus the number of intervals, $m$. 
  Different values of $l$ are considered (ranging from $6$ up to $12$). 
  Data refers to: 
  $S_0=100$, $r=9$ \%, $\delta_t=t_i-t_{i-1}=1/12$ year, $\sigma = 30$ \%, 
  $L=0$, $H=m \, h$ with $h= 4$ \% and $m=$ ranging from $4$ up to to $40$. 
  }
}
\label{Fig:reverse_cliquet_PME_versus_QMC}
\end{center}
\end{figure}
%
%
In figure~\ref{Fig:reverse_cliquet_PME_versus_QMC} we plot relative error 
between PME results ($l$, ranging from $6$ up to $12$) and QMC for 
different values of $m$, keeping the time interval constant. As we have 
stated in previous paragraph, the PDF of sum of negative performances 
converges, in the limit $m \rightarrow \infty$, to a gaussian distribution, 
making PME asymptotically exact. As a consequence the agreement between 
PME and QMC becomes better and better increasing the number of 
intervals $m$.

%
\subsection{Barrier options with discrete monitoring dates} 
\label{subsection:barrier_option_pricing}

\subsubsection{Barrier options: contract description} 
\label{subsubsection:barrier_contract_des}

The pay-off of a generic knock out barrier option can be written 
as follows: 
\begin{equation}
\mbox{Pay-off}_{\mbox{{\tiny Barrier}}} = \left\{ \begin{array}{ll}
{\cal F}[S(T)] & \textrm{if $S(t_i) > B_i$ for all observation dates 
$\{ t_i \}$} \\
b 	& \textrm{otherwise}
\end{array} \right.
\label{eq:pay_off_barrier_opt}
\end{equation}
where: ${\cal F}$ is a generic positive function which represents the 
option pay-off if the barrier has not been touched, $b$ is known as 
``rebate'' and represents the amount paid by the contract if the barrier 
has been touched (often $b=0$), $B_i$ is the barrier level at time 
$t_i$\footnote{For constant barrier level $B_i=B \quad \forall i$, in 
the present discussion we do not make any assumption about $\{B_i\}$.}, 
$\{t_i \}_{i=1, 2, ..... N}$ are the discrete observation dates set, $S_0$ 
the spot price, $S(t_i)$ the underlying price at time $t_i$. 
%

\subsubsection{Barrier options: PME problem formulation} 
\label{subsubsection:barrier_pme_formulation}

The problem implicit in a barrier option can be reformulated
in terms of PDF's. The stock price a time $t_i$ is given by: 
\begin{equation}
S(t_i) = S_0 \, e^{\sum_{j=1}^{i} \left ( r - \frac{\sigma^2}{2} \right ) \, 
\delta t_j + \sigma \, \sqrt{\delta t_j} \, e_j  } \, ,
\label{eq:barrier_opt_S_i}
\end{equation}
where $\delta t_j = t_j - t_{j-1}$ and $\{e_j\}$ are independent 
\underline{normally} distributed random variables with zero mean and 
unit variance. \\
Let us define the following stochastic variables:
\begin{equation}
z_i = \left ( \log \frac{S(t_i)}{S_0} \mid S(t_j)>B_j 
\quad \forall j \le i \right ) \, ,
\label{eq:barrier_opt_def_z_i} 
\end{equation}
which embedding the condition that all observations until time 
$t_i$ are above the barrier level. \\
Let us introduce:
\begin{equation}
p_i = \mbox{probability that:  } S(t_j)>B_j \qquad \forall j \le i \; ,
\label{eq:barrier_opt_def_p_i}
\end{equation}
i.e. the probability that the option is alive at time $t_i$. \\
Then, the option price can be expressed in terms of PDF of $z_m$, 
$P_{z_m}$, as:
\begin{equation}
\mbox{option price}_{\mbox{{\tiny KO}}} = e^{-r \, (t_m-t_0) } \left[ 
(1-p_m) \, b + p_m \, \int {\cal F}(S_0 \, e^{x}) \, P_{z_m}(x) \, dx
\right]  \; .
\label{eq:barrier_opt_price_via_pdf}
\end{equation}
The stochastic variables $\{z_i\}$ can be recursively related one to 
each other, by the following equations: 
\begin{eqnarray}
&& z_0 = 0 \, , \nonumber \\
&& w_i =  z_{i-1} + \left ( r - \frac{\sigma^2}{2} \right ) \, \delta t_i \,
+ \sigma \, \sqrt{\delta t_i} \, e_i  \, , 
\label{eq:barrier_opt_recursive_eq_for_z_i} \\
&& z_i = \left ( w_i \mid S(t_i) > B_i \right ) 
= \left ( w_i \mid w_i > \log \frac{B_i}{S_0} \right ) \, , \nonumber 
\end{eqnarray}
where we have introduced the dummy variables $w_i$. \\
The equations set~(\ref{eq:barrier_opt_recursive_eq_for_z_i}) can be 
translated in terms of PDF's as follows:
\begin{eqnarray}
&& P_{z_0}(x) = \delta(x) \, ,  
\label{eq:barrier_opt_recursive_eq_for_pdf_1} \\
&& P_{w_i} = P_{z_{i-1}} \circ \Phi_{(r-\sigma^2 / 2) \delta t_{i},
\sigma^2 \, \delta t_i } \, , 
\label{eq:barrier_opt_recursive_eq_for_pdf_2} \\
&& P_{z_i}(x) = 
		\left\{ \begin{array}{ll}
		\frac{1}{k_i} \, P_{w_i}(x) & \textrm{if 
		$x>\log \frac{B_i}{S_0}$} \\
		0 & \textrm{otherwise}
		\end{array} \right.
\label{eq:barrier_opt_recursive_eq_for_pdf_3} 
\end{eqnarray}
where: $\delta(x)$ is the Dirac delta distribution, ``$\circ$'' indicates 
the convolution product, $\Phi_{\rho,v^2}$ is the normal distribution 
with mean $\rho$ and variance $v^2$ and $k_i$ is the probability to 
find $w_i$ below $\log \frac{B_i}{S_0}$:
\begin{equation}
k_i = \int_{\log \frac{B_i}{S_0}}^{\infty} P_{w_i}(x) \, dx  \, .
\label{eq:barrier_opt_k_i_def}
\end{equation}
In this framework, the probabilities $p_i$ are then related by the
recursive relation:
\begin{equation}
p_i = p_{i-1} \cdot k_i \, .
\label{eq:barrier_opt_p_i_recursive_relation}
\end{equation}
The PME iterative scheme for a barrier option evaluation can be therefore 
structured as follows:
\begin{itemize}

	\item[(A)] 
	fix the number of moments, $l$, to be used in PME algorithm;

	\item[(B)]
	repeat, starting from $i=0$, the steps B1--B3 until $i=m$ is 
	reached:
	 
	\begin{itemize}
		\item[(B1)]
		if $i=0$, $P_{z_{0}}$ is the Dirac delta distribution, 
		therefore its moments, at any order are identically zero 
		apart from moment of order zero which is, by definition, 
		equal to 1. \\
		Otherwise for $i>0$: knowing the moments of $P_{z_{i-1}}$, 
		compute all moments, up to order $l$, of PDF $P_{w_i}$ by 
		using equation~(\ref{eq:moment_composition_law})\footnote{The
		moments of a gaussian distribution are known at every 
		order, see equation~(\ref{eq:gaussian_moments}).}.

		\item[(B2)]
		by reverting to equations~(\ref{eq:pdf_decomposition}), 
		(\ref{eq:gram_charlier_series}) 
		and~(\ref{eq:linear_system_eq_for_moments}), reconstruct 
		PDF $P_{w_i}$ by a Gram Charlier series around normal 
		density function $\Phi_{0,1}$.

		\item[(B3)]
		given $P_{w_i}$, from 
		eq.~(\ref{eq:barrier_opt_recursive_eq_for_pdf_3}), compute 
		all moments of $P_{z_{i}}$ up to order $l$.
	\end{itemize}

	\item[(C)]
	Reached $i=m$, we know all $P_{z_m}$ moments up to $l$, again, 
	by using equations~(\ref{eq:pdf_decomposition}), 
	(\ref{eq:gram_charlier_series}) 
	and~(\ref{eq:linear_system_eq_for_moments}) we can develop 
	$P_{z_m}$ as a Gram Charlier series around normal density. 
	One that is done, the option price can be easily evaluate by 
	means of eq.~(\ref{eq:barrier_opt_price_via_pdf}).

\end{itemize}

\begin{figure}[hbtp]
\begin{center}
\epsfig{file=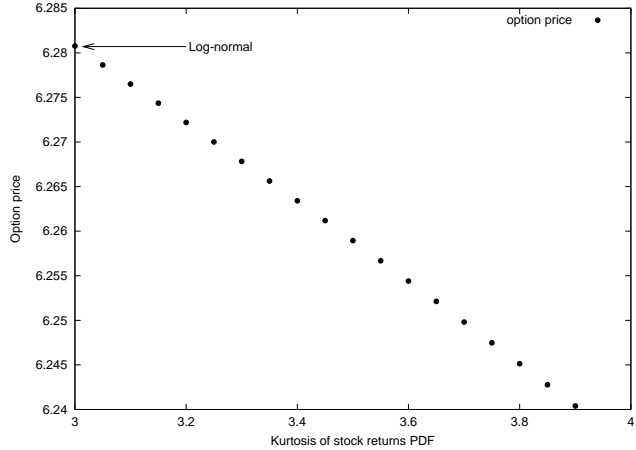,scale=0.34,angle=-90}
\caption{
	{ \small
	Price of a down-out call option, calculated with PME ($l=20$) 
	by considering a leptokurtic  behavior (i.e. kurtosis grater 
	than 3) in probability density function of  stock returns.
	$S_0=100$, $r=10$ \%, $T=0.2$ year, $E=100$, $\sigma = 30$ \%, 
	$B=89$ and $m=5$. 
	}
}
\label{Fig:barrier_opt_price_versus_kurtosis}
\end{center}
\end{figure}

It interestingly to note that within the scheme presented above, we need
to evaluate only integrals of the form:
\begin{equation}
I_n(d) = \frac{1}{\sqrt{2 \, \pi}} \, \int_{-d}^{+\infty} \, x^n \, 
e^{-\frac{1}{2} x^2} \, ,
\label{eq:barrier_opt_integrals_to_be_solved}
\end{equation}
which can be reduced to the evaluation of 
the well known inverse cumulative normal function\footnote{For instance:
\begin{eqnarray}
&& I_0(d) =  \mbox{Err\_f}(d) \, ,  \nonumber \\
&& I_1(d) = \frac{1}{\sqrt{2 \, \pi}} \, e^{-\frac{1}{2} d^2} \, ,  \nonumber \\
&& I_2(d) = \mbox{Err\_f}(d) - d \, \frac{1}{\sqrt{2 \, \pi}} \, e^{-\frac{1}{2} d^2} \, ,  \nonumber \\
&& I_3(d) = (2 \, +d^2) \, \frac{1}{\sqrt{2 \, \pi}} \, e^{-\frac{1}{2} d^2} \, ,  \nonumber \\
&& .  \nonumber \\
&& .  \nonumber 
\end{eqnarray}
}. 
Therefore we do not need any numerical methods to compute the required 
integrals. \\
A last remark on PME scheme: if we want to treat also non log-normal 
processes for equity prices evolution, the only modification to be 
implemented regards the equations~(\ref{eq:barrier_opt_recursive_eq_for_z_i}) 
and~(\ref{eq:barrier_opt_recursive_eq_for_pdf_2}), where in place of $e_i$ 
and $\Phi_{0,1}$ we must consider a non normal stochastic variable/PDF.
Hence, when computing the $P_{w_i}$ moments via 
eq.~(\ref{eq:moment_composition_law}), we just need to consider 
the real PDF moments, different from those of a normal distribution.
As an example in figure~\ref{Fig:barrier_opt_price_versus_kurtosis}, 
we report how the option price changes when fat tails (i.e. non-normal 
kurtosis) are considered.  
%
%
%
%
%
%
%
%
%
As expected, increasing leptokurticity of stock returns distribution has 
the effect  to enlarger the chances of touch the barrier before maturity, 
lowering therefore the option value. 


\subsubsection{Barrier options: PME option pricing, a comparison with 
other techniques} 
\label{subsubsection:barrier_numerical_results}

In order to make a quantitative comparison with other numerical methods, 
we have considered a down-out barrier option with constant barrier $B$. 
The pay-off at maturity is given by:
\begin{equation}
\mbox{Pay-off} = 
\left\{ \begin{array}{ll}
S-E & \textrm{if $S > E$ }\\
0 	& \textrm{otherwise}
\end{array} \right.
\label{eq:pay_off_barrier_down_out_call_option}
\end{equation}
i.e. the usual call plain vanilla pay-off. \\
For this kind of option, recently Fusai et 
al.~\cite{Fusai_analytical_result_barrier_opt} have derived a closed 
form solution. 
Table~\ref{Tab:barrier_PME_versus_other} shows, for different values of 
barrier level, $B$, and observation dates number, $m$, a comparison of 
PME with the exact result and other numerical techniques.

\newpage 
%
%
%
%
{
 
\begin{table}[h]
\tiny

\begin{center}
\begin{tabular}{|l|l|l|l|l|l|l|l|l|}
\hline
\multicolumn{9}{|c|} {} \\
\multicolumn{9}{|c|}{\bf \large Barrier option pricing with different techniques.} \\
\multicolumn{9}{|c|} {} \\
\hline \hline
{\bf B} & 89          &  95         &  97         &  99         &  89         &  95         &  97         &  99         \\
        &             &             &             &             &             &             &             &             \\
\hline
{\bf m} &  5          &  5          &  5          &  5          &  25         &  25         &  25         &  25         \\
        &             &             &             &             &             &             &             &             \\
\hline \hline
        &             &             &             &             &             &             &             &             \\
PME(4)  & $6.27407$   & $5.68309$   & $5.17957$   & $4.49928$   & $6.20762$   & $5.11871$   & $4.14616$   & $2.82893$   \\
        &             &             &             &             &             &             &             &             \\
\hline
        &             &             &             &             &             &             &             &             \\
PME(12) & $6.27954$   & $5.67106$   & $5.16749$   & $4.48947$   & $6.19971$   & $5.08100$   & $4.11596$   & $2.81157$   \\
        &             &             &             &             &             &             &             &             \\
\hline
        &             &             &             &             &             &             &             &             \\
PME(20) & $6.28077$   & $5.67109$   & $5.16726$   & $4.48918$   & $6.20850$   & $5.08032$   & $4.11508$   & $2.81154$   \\
        &             &             &             &             &             &             &             &             \\
\hline
        &             &             &             &             &             &             &             &             \\
PME(32) & $6.28076$   & $5.67110$   & $5.16725$   & $4.48917$   & $6.21032$   & $5.08096$   & $4.11559$   & $2.81224$   \\
        &             &             &             &             &             &             &             &             \\
\hline
        &             &             &             &             &             &             &             &             \\
Exact   & $6.28076$   & $5.67111$   & $5.16725$   & $4.48917$   & $6.20995$   & $5.08142$   & $4.11582$   & $2.81244$   \\
        &             &             &             &             &             &             &             &             \\
\hline
        &             &             &             &             &             &             &             &             \\
QMC     & $6.28075$   & $5.67111$   & $5.16726$   & $4.48912$   & $6.210005$  & $5.08156$   & $4.11561$   & $2.81233$   \\
        &             &             &             &             &             &             &             &             \\
\hline
        &             &             &             &             &             &             &             &             \\
NRI     & $6.2763 $   & $5.6667 $   & $5.1628 $   & $4.4848 $   & $6.2003 $   & $5.0719 $   & $4.1064 $   & $2.8036 $   \\
        &             &             &             &             &             &             &             &             \\
\hline
        &             &             &             &             &             &             &             &             \\
CMF     & $6.284  $   & $5.646  $   & $5.028  $   & $4.050  $   & $6.210  $   & $5.084  $   & $4.113  $   & $2.673  $   \\
        &             &             &             &             &             &             &             &             \\
\hline
        &             &             &             &             &             &             &             &             \\
TT      & $6.281  $   & $5.671  $   & $5.167  $   & $4.489  $   & $6.21   $   & $5.081  $   & $4.115  $   & $2.812  $   \\
        &             &             &             &             &             &             &             &             \\
\hline
        &             &             &             &             &             &             &             &             \\
SRQM    & $6.2809 $   & $5.6712 $   & $5.1675 $   & $4.4894 $   & $6.2101 $   & $5.0815 $   & $4.11598$   & $2.8128 $   \\
        &             &             &             &             &             &             &             &             \\
\hline
\end{tabular}
\end{center}
\caption{
	{\small
	Option pricing for a barrier option with: 
	$S_0=100$, $E=100$, $r=10$ \%, $T=0.2$ year, $\sigma = 30$ \%. 
	The table shows option price computed with different techniques: 
	PME ($l=4$, $12$, $20$ and $32$), 
	Exact results due to Fusai~\cite{Fusai_analytical_result_barrier_opt},
	QMC - Quasi Monte Carlo method (calculated with $2^{27}-1$ scenarios), 
	NRI - Numerical Recursive Integration (results are due to Aitsahlia~\cite{Aitsahlia_quadrature_barrier_opt}), 
	CMF - Continuous Monitoring Formula (Broadie in~\cite{Broadie_tree_barrier_opt}),
	TT - trinomial tree (results are due to Broadie~\cite{Broadie2_tree_barrier_opt}) and
	SRQM - Simpson Recursive Quadrature Method (Fusai in~\cite{Fusai_quadrature_barrier_opt}). 
	}
}
\label{Tab:barrier_PME_versus_other}
\end{table}
}
%
%

\begin{figure}[hbtp]
\begin{center}
\epsfig{file=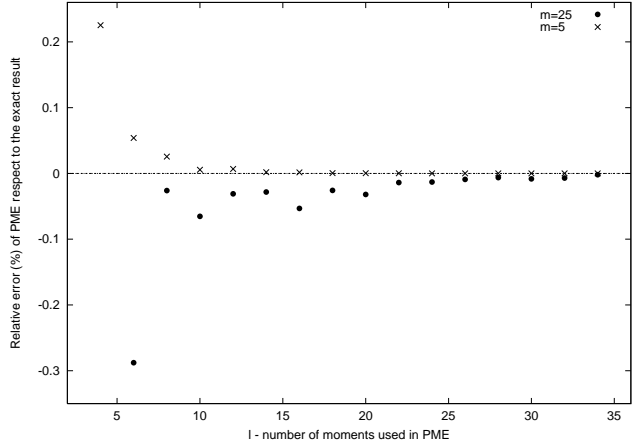,scale=0.34,angle=-90}
\caption{
  { \small
  Percentage error between PME estimates ($l=4$, $6$, ...., $34$) and
  exact result for a down-out call option with 
  $S_0=100$, $r=10$ \%, $T=0.2$ year, $E=100$, $\sigma = 30$ \%, $B=99$ 
  and $m=5$, $25$. 
  }
}
\label{Fig:barrier_opt_PME_versus_EXACT}
\end{center}
\end{figure}

The agreement is quite good and it is better, in general, than the 
corresponding result achievable by simple NRI, at least when the number 
of observation dates is relatively small (as it is in the example considered 
in table~\ref{Tab:barrier_PME_versus_other}). \\
In figure~\ref{Fig:barrier_opt_PME_versus_EXACT}, we report the relative 
percentage error between PME estimates and the exact result, for two 
different monitoring frequencies. The graph shows the accuracy of the PME 
method, which is already good (errors less than 0.1 \%) just including few 
moments. \\
%
%
%
%
%
%
%
%
%
However the number of moments to be considered in order to reach a 
good precision is higher respect to asian options (see 
chapter~\ref{subsection:asian_option_pricing}) or reverse cliquet 
options (see chapter~\ref{subsection:reverse_cliquet_option_pricing}). 
The reason being the strong skewness (especially when equity spot price 
is near to the barrier or $m$ is large) characterizing the underlying 
probability distribution, $P_{z_m}$, which leads to a strong departure 
from gaussian behavior. Indeed large skewness values lead to a 
Gram-Charlier expansions which may not be non-negative, especially on 
the distribution tails (compromising therefore the acceptance of the 
function as a true density). Because in a barrier option a key point 
is to measure the probability to touch the barrier at each step 
(i.e. to calculate accurately an integral on the left tail) the non 
positive definiteness of the PDF's tails, may lead to errors that 
compromise the pricing quality. That requires, especially when $m$ 
is large, a correction to simple PME scheme, in order to fix (at least
partially) the problem. A simple trick consists to substitute the 
step (B2) of the algorithm described in the previous paragraph, by a 
two layer procedure:

\begin{itemize}
	\item[(B2)] 

	\begin{itemize}
		\item[]

		\item[(L1)] reconstruct PDF, $P_{w_i}$, as a perturbative 
		series expansion around normal density:
		$
		P_{w_i}= \left ( \sum_{i=0}^{K} c_i \, x^i \, \right ) \, 
		\Phi_{0,1}(x) \; 
		$;
	
		\item[(L2)] (a) find the lower negative value $x_0$ such 
		that: $ \sum_{i=0}^{K} c_i \, x_{0}^{i} = 0$; \\
		(b) considering a new function
		$
		\hat P_{w_i}(x) = 
		\left\{ \begin{array}{ll}
		P_{w_i}(x) & \textrm{if $x \ge x_0$ }\\
		0 	& \textrm{otherwise}
		\end{array} \right.
		$
		\\
		Due to the truncation procedure, we can expect that 
		$\hat P_{w_i}$  moments depart slightly from initial 
		target moments. 
		However we can apply again the perturbative expansion 
		procedure using as basic PDF the $\hat P_{w_i}$ 
		function~\footnote{Before that, it is necessary to 
		rescale $\hat P_{w_i}$, according to 
		eq.~(\ref{eq:pdf_decomposition}), in order to obtain a 
		PDF with zero mean and unit variance.} instead of a 
		simple gaussian. 

	\end{itemize}

\end{itemize}

In that way it is possible to derive an improved perturbative series 
expansion, where the problem of non positiveness is partially 
fixed~\footnote{Indeed, the above procedure does not guarantee that 
the new expansion is non negative for any value of $x$. However numerical 
evidence shows a good improvement and a significantly reduction of the 
non positiveness problem.}. \\
In table~\ref{Tab:barrier_PME_versus_EXACT}, we report PME valuations 
(using such improved algorithm and including the first twelve moments, 
$ l=12 $) versus exact results. The error is typically lower than 
$0.1$ \%, also for large number of observation dates, $m$. 
%
%
%
{
 
\begin{table}[h]
\tiny

\begin{center}

\begin{tabular}{|l|l|l|l|}
\hline
\multicolumn{4}{|c|} {} \\
\multicolumn{4}{|c|}{\bf \large Barrier options pricing } \\
\multicolumn{4}{|c|} {\bf \small (improved PME / exact results)}\\
\multicolumn{4}{|c|} {} \\
\hline \hline
                        &                     &                     &                     \\
 m                      & PME(12)             &  Exact result       & Relative error      \\
                        &                     &                     &                     \\
\hline \hline
                        &                     &                     &                     \\
10                      &  $4.1820$           &  $4.18224$          &  $0.006$  \%        \\
                        &                     &                     &                     \\
50                      &  $3.1260$           &  $3.12633$          &  $0.011$  \%        \\
                        &                     &                     &                     \\
80                      &  $2.9391$           &  $2.93918$          &  $0.003$  \%        \\
                        &                     &                     &                     \\
100                     &  $2.8643$           &  $2.86442$          &  $0.004$  \%        \\
                        &                     &                     &                     \\
120                     &  $2.8087$           &  $2.80903$          &  $0.012$  \%        \\
                        &                     &                     &                     \\
150                     &  $2.7461$           &  $2.7474$           &  $0.047$  \%        \\
                        &                     &                     &                     \\
180                     &  $2.7020$           &  $2.70163$          &  $-0.014$ \%        \\
                        &                     &                     &                     \\
200                     &  $2.6746$           &  $2.67682$          &  $0.083$  \%        \\
                        &                     &                     &                     \\
220                     &  $2.6531$           &  $2.65545$          &  $0.089$  \%        \\
                        &                     &                     &                     \\
250                     &  $2.6253$           &  $2.628099$         &  $0.107$  \%        \\
                        &                     &                     &                     \\
280                     &  $2.6018$           &  $2.60534$          &  $0.136$  \%        \\
                        &                     &                     &                     \\
300                     &  $2.5879$           &  $2.59056$          &  $0.103$  \%        \\
                        &                     &                     &                     \\
500                     &  $2.5018$           &  $2.50259$          &  $0.032$  \%        \\
                        &                     &                     &                     \\
\hline
\end{tabular}
\end{center}
\caption{
	{\small
	Option pricing for a barrier option with: 
	spot price = 100, strike = 100, barrier = 98, $r=10$ \%, 
	$T=0.2$ year, $\sigma = 30$ \%. 
	The table shows a comparison between the option value computed 
	with an improved PME technique ($l=12$) against the exact result 
	(Fusai et al. in~\cite{Fusai_analytical_result_barrier_opt}) for
	different $m$ values. 
	}
}
\label{Tab:barrier_PME_versus_EXACT}
\end{table}
}
%
%

%
%
\section{Summary and Conclusions} 
\label{section:conclusions}

In this article we have proposed a new methodology for option pricing
belonging to quadrature techniques. 
Respect to other quadrature algorithms, where for instance density 
functions are modeled through a grid of points or polynomial 
interpolations, our method adapts a Gram–Charlier series expansion 
around a given distribution (usually a gaussian function). \\
The highlights features of the method are the following: 
\begin{itemize}
	\item[(i)]
	by representing PDF's through their moments, we are able to 
	capture all the essential features of the problem by using 
	just few parameters, improving computational performances and
	memory usage. How we have shown in the paper, the 
	crude choice of retaining only the first four moments (i.e. 
	kurtosis and skewness) in PME scheme, provides an excellent 
	approximation of the option value (with errors less than $0.1$\% 
	for asian and reverse cliquet options). Furthermore, by 
	increasing the number of moments retained in PDF expansion, the 
	numerical solution converges to the exact result, making possible, 
	in principle, to increase progressively the estimate precision. 

	\item[(ii)]
	The method is extremely simple and natural from a conceptual 
	point of view and therefore easily implementable.

	\item[(iii)]
	It is extensible to different pay-off contracts (indeed the 
	programming code needs very few modifications changing the 
	contract features).

	\item[(iv)]
	non log-normal processes (i.e. stochastic processes 
	characterized by a non log-normal PDF), can be naturally treated 
	within a PME scheme, without any modification to the algorithm. 

\end{itemize}
Moreover, the proposed method, maintains all the typical advantages
of quadrature methods, that is:
\begin{itemize}
	\item[(i)]
	we need to perform computations only at trigger times;

	\item[(ii)]
	the CPU time scales linearly with the number of observation 
	dates.

	\item[(iii)]
	the price estimate is not too sensitive to changes in volatility;

	\item[(iv)]
	there are no time discretization errors.

	\item[(v)]
	it is easily to incorporate additional exoticity in the contract
	(e.g. for a barrier option, it is straightforward to include a 
	time varying barrier).

\end{itemize}


%
%

%
%

\newpage

%
%


\begin{thebibliography}{99}

%
%

\bibitem{Gradshteyn} 
Gradshteyn, I.S. and Ryzhnik, I.M., 1994. Table of Integrals, Series, and 
Products. (Academic Press, 5th edition).

%
%





%
%

\bibitem{Black} 
F. Black and M. Scholes, {\it The Pricing of Options and Corporate 
Liabilities}, Journal of Political Economics, \textbf{81}, 637--654 (1973).

\bibitem{Fusai_analytical_result_barrier_opt} 
G. Fusai, I. D. Abrahams and C. Sgarra, {\it An Exact Analytical Solution 
For Discrete Barrier Options}, preprint. 


%
%

\bibitem{Hull_tree}
J. C. Hull and A. White, {\it Efficient procedures for valuing European 
and American path-dependent options}, Journal of Derivatives, 
\textbf{1}, 21--31 (1993).

\bibitem{Barraquand_tree}
J. Barraquand and T. Pudet, {\it Pricing of American path-dependent 
contingent claims}. Mathematical Finance, \textbf{6}, 17--51 (1996).

\bibitem{Steiner} 
M. Steiner, M. Wallmeier and R. Hafner, {\it Pricing near the barrier: 
the case of discrete knock-out options}, Journal of Computational Finance, 
vol. \textbf{3, n. 1, fall}, 69--90 (1999).

\bibitem{Boyle_tree}
P. P. Boyle and S.H. Lau, {\it Bumping up against the barrier with 
the binomial method}, Journal of Derivatives, \textbf{1}, 6--14 (1994).

\bibitem{Broadie_tree_barrier_opt}
M. Broadie, P. Glasserman and S. Kou, {\it Connecting Discrete and 
Continuous Path-Dependent Options}, Finance and Stochastic, 
\textbf{3}, 55--82 (1999).

\bibitem{Broadie2_tree_barrier_opt}
M. Broadie, P. Glasserman and S. Kou, {\it A Continuity Correction 
for Discrete Barrier Options}, Mathematical Finance, 
\textbf{7}, 325--349 (1997).



%
%
\bibitem{Kemna_monte_carlo_asian_opt}
A. G. Z. Kemna and A. C. F. Vorst, {\it A pricing method for options 
based on average asset values}, Journal of Banking and Finance, 
\textbf{14}, 113--129 (1990).

\bibitem{Boyle}
P. Boyle, M. Broadie and P. Glasserman, {\it Monte Carlo Methods for 
Security Pricing}, Journal of Economic Dynamics and Control, 
\textbf{21}, 1267--1321 (1997).

\bibitem{Jackel}
P. Jackel, {\it Monte Carlo Methods in Finance}, 
Wiley Finance, (2002).


%
%

\bibitem{Wilmott_pde_1}
P. Wilmott, J.N. Dewynne and S. Howison, {\it Option Pricing: 
Mathematical Models and Computation}, Oxford Financial Press. 
(1993)

\bibitem{Wilmott_pde_2}
J. N. Dewynne and P. Wilmott, {\it A note on average rate options 
with discrete sampling}. SIAM Journal on Applied Mathematics, 
\textbf{55}, 267–-276 (1995).

\bibitem{Tavella_pde}
D. Tavella and C. Randall, {\it Pricing Financial Instruments: 
The Finite Difference Method}, Wiley Series in Financial 
Engineering, New York: John Wiley \& Sons (2000).



%
%


\bibitem{Aitsahlia_quadrature_barrier_opt}
F. Aitsahlia and T. L. Lai, {\it Valuation of Discrete Barrier and 
Hindsight Options}, The Journal of Financial Engineering, 
vol. \textbf{6}, n. \textbf{2}, 169--77 (1997).

\bibitem{Aitsahlia_quadrature_lookback_opt}
F. Aitsahlia and T. L. Lai, {\it Random Walk Duality and the Valuation of 
Discrete Look-back Options}, Applied Mathematical Finance, 
vol. \textbf{5}, n. \textbf{3}, 227--240 (1998).

\bibitem{Sullivan_quadrature_barrier_opt}
M. A. Sullivan, {\it Pricing discretely monitored barrier options, Journal 
of Computational Finance}, vol. \textbf{3}, n. \textbf{4}, summer, 
35--52 (2000).

\bibitem{Lim_quadrature_asian_opt}
T. W. Lim, {\it Performance of recursive integration for pricing
European-style Asian options}, preprint. 

\bibitem{Andricopoulos_quadrature_methods}
A. D. Andricopoulos, M. Widdicks, P. W. Duck, and D. P. Newton, {\it Universal
Option Valuation Using Quadrature Methods}, Journal of Financial Economics, 
\textbf{67}, 447--471 (2003).

\bibitem{Fusai_quadrature_barrier_opt} 
G. Fusai and M. C. Recchioni, {\it Analysis of Quadrature Methods for 
Pricing Discrete Barrier Options}, preprint.  \\
G. Fusai and M. C. Recchioni, {\it Numerical Valuation of Discrete Barrier 
Options}, Financial Options Research Center Preprint, 2001/119, Warwick 
Business School (2001).

%
%


\bibitem{Gram-Charlier_series}
C. V. L. Charlier, {\it \"Uber das Fehlergesetz}, 
Ark. Math. Astr. och Phys. 2, \textbf{8}, 1--9, (1905--06). \\
P. L. Chebyshev, {\it Sur deux th\'eor\`emes relatifs aux probabilit\'es}, 
Acta Math., \textbf{14}, 305--315, (1890). \\
F. Y. Edgeworth, {\it The Law of Error}, Cambridge Philos. Soc.,
\textbf{20}, 36--66 and 113--141, (1905). 

\bibitem{Cramer}
H. Cram\'er, {\it On Some Classes of Series Used in Mathematical 
Statistics}, Proceedings of the Sixth Scandinavian Congress of 
Mathematicians, Copenhagen, 399--425, (1925). \\ 
D. L. Wallace, {\it Asymptotic Approximations to Distributions}, 
Ann. Math. Stat., \textbf{29}, 635--654, (1958). \\ 
G. Szego, {\it Orthogonal Polynomials}, 4th ed. Providence, 
RI: Amer. Math. Soc., (1975). 

\bibitem{Kolmogorov}
B.V. Gnedenko and A.N. Kolmogorov, {\it Limit distribution for sums of 
independent random variables}, Addison-Wesley, cambridge MA, (1954).

\bibitem{Jarrow}
R. Jarrow, A. Rudd, {\it Approximate option valuation for arbitrary 
stochastic processes}, Journal of Financial Economics, \textbf{10},
347--369, (1982).

\bibitem{Corrado_gram}
C. J. Corrado and Tie Su, {\it Implied volatility skews and stock 
return skewness and kurtosis implied by stock option prices}, 
The European Journal of Finance, \textbf{3}, 73-–85 (1997). 

\bibitem{Knight_gram}
J. Knight, S. Satchell, {\it Pricing derivatives written on assets 
with arbitrary skewness and kurtosis}, Trinity College, mimeo (1997).




\end{thebibliography}
\end{document}